\documentstyle[12pt,epsfig]{article}
\def\beq{\begin{equation}}   \def\eeq{\end{equation}}
\def\bea{\begin{eqnarray}}   \def\eea{\end{eqnarray}}

\newcommand{\gsim}{\lower.7ex\hbox{$
\;\stackrel{\textstyle>}{\sim}\;$}}
\newcommand{\lsim}{\lower.7ex\hbox{$
\;\stackrel{\textstyle<}{\sim}\;$}}

\newcommand{\La}{\overline{\Lambda}}

\newcommand{\bibit}[1]{\bibitem{#1}}

\newcommand{\matel}[3]{\langle #1|#2|#3\rangle}
\newcommand{\aver}[1]{\langle #1\rangle}

\newcommand{\state}[1]{ |#1\rangle}

\begin{document}
\def\lsim{\mathrel{\rlap{\lower3pt\hbox{\hskip0pt$\sim$}}
    \raise1pt\hbox{$<$}}}         
\def\gsim{\mathrel{\rlap{\lower4pt\hbox{\hskip1pt$\sim$}}
    \raise1pt\hbox{$>$}}}         

\begin{flushright}
UND-HEP-00-BIG\hspace*{.2em}03\\
hep-ph/0005278\\
\end{flushright}
\vspace{1.25cm}
\begin{center} \Large 
{\bf Analytical Heavy Quark Expansion in the \\
't~Hooft Model}\\
\end{center}
\vspace*{.8cm}
\begin{center} {\Large 
Matthias Burkardt$^{\,a}$ \,{\large and} \,Nikolai Uraltsev$^{\,b,c}$}
\vspace*{.98cm}\\
{\normalsize 
$^a${\it Dept.\ of Physics,
New Mexico State Univ.,
Las Cruces, NM 88003-0001, U.S.A.}
\\
$^b${\it Physics Dept.,
Univ. of Notre Dame du
Lac, Notre Dame, IN 46556, U.S.A.} 
\\
$^c${\it Petersburg Nuclear Physics Inst., Gatchina, 
St.\,Petersburg, 188350, Russia}} \\
\vspace{.3cm}
\vspace*{2.94cm}

{\Large{\bf Abstract}}\\
\end{center}

We present a number of exact relations for the heavy quark limit and
develop an analytical $1/m_Q$ expansion for heavy mesons in the
't~Hooft model. Among the new results are relation
$3\mu_\pi^2=\La^2-m_{\rm sp}^2+\beta^2$, $\;1/m_Q$ corrections to the
decay constants, to the kinetic expectation values and $1/m_Q^2$
nonperturbative corrections to the $B\to D$ amplitude at zero
recoil. The properties of the IW functions are addressed and the
small velocity sum rules are verified.

\thispagestyle{empty}
\addtocounter{page}{-1}

\newpage

\section{Introduction}

Heavy quark symmetry and the heavy quark expansion have played an 
important role in understanding weak decays of charm and beauty
hadrons allowing extraction of the fundamental parameters of the 
Standard Model. Beauty and, in particular, charm quarks are not
infinitely heavy even in the crude approximation. In practice,
$1/m_Q$ corrections to the strict $m_Q\to \infty$ limit often
constitute the main limitation. Even in a few cases where the first
terms are known, a question remains about the convergence of the
employed $1/m_Q$ expansion. (For a review and further references, see 
Ref.\,\cite{rev}.)

In such a situation it is advantageous to have a model laboratory
where both $1/m_Q$ corrections, and the whole finite-$m_Q$ amplitudes 
can be evaluated exactly. On the one hand, this allows to trace in
detail how the methods used for actual QCD work in a simplified
setting. On the other hand, it has been empirically observed that
certain quantities in the heavy flavor hadrons suffer from
numerically large power corrections, whilst others seem robust
against finite mass effects. Studying this in the toy models can help
gaining some insights which can be applied in various studies of
actual QCD, including those based on the QCD sum rules technique or
lattice computations.

One such solvable model, which has been applied to 
a variety of strong interaction phenomena in the past is 
the 't~Hooft model,  QCD in 1+1 dimensions at large number of colors 
$N_c\to \infty$ \cite{thooft}.
The 't~Hooft model has two important features that resemble
real QCD$_{3+1}$ phenomenologically: confinement and spontaneous 
chiral symmetry breaking. Since the underlying microscopic
mechanisms for these effects in this model are quite different
from the respective mechanisms in QCD$_{3+1}$, it is probable
that QCD$_{1+1}$ is of little
help to understand the {\it origin} of these features in real 
QCD, but it may nevertheless be very valuable to better 
understand the {\it consequences} for other observables.

In this paper we will focus on the 
$m_Q \to \infty$ limit of the 't~Hooft model.
Even though we were not able to  solve this limit analytically,
we succeeded to derive a number of exact relations involving
terms that appear in the $1/m_Q$ expansion. 
In Section 2 we will discuss some general features of this 
model. In Section 3 the static limit  $m_Q\rightarrow \infty$
will be explored and we will derive a number of relations
involving 't~Hooft wavefunctions in this limit. 
These relations will be applied in Section 4 to derive some
exact results for power  corrections in this model. Section 5 gives
the conclusions.

\section{The 't~Hooft Model}

QCD$_{1+1}$ is based on the Lagrangian 
\beq
{\cal L} = \sum_q \bar{q}\left(
i\gamma^\mu D_\mu -m_q\right)q \:-\: \frac{1}{2g_s^2}\, 
\mbox{Tr} \: G^{\mu \nu} G_{\mu \nu}\; ,
\eeq
defined in one space and one time dimension, i.e.
$\mu,\nu=0,1$ and all fields depend only on two space-time
coordinates.
In the light-cone gauge, $A^+=0$, as in any other
axial gauge, the non-Abelian term in the field strength
tensor 
$G_{\mu \nu} = i\left[D_\mu, D_\nu\right]=
\partial_\mu A_\nu -\partial_\nu A_\mu
+ i \left[A_\mu, A_\nu\right]$ 
vanishes, which drastically
simplifies the dynamics of the model. 
This allows
one to eliminate the only non-vanishing component of the
gauge field $A^-$, by means of the Poisson equation 
\beq
-\partial_-^2 A^-_a = g_s^2 J^+_a\; ,
\label{eq:poisson}
\eeq
where we have introduced light-cone coordinates
$x^\pm \!=\! \frac{1}{\sqrt{2}}\left(x^0\!\pm \!x^1\right)\,$.
In two dimensions there are no dynamical gluons and,
after solving the Poisson equation (\ref{eq:poisson}),
the only remnant of the gluon field is a Coulomb like
instantaneous interaction among the quarks. Of course, 
in one space dimension, the `Coulomb' interaction is 
linearly confining.
It is this feature which makes QCD$_{1+1}$ so attractive
if one is interested in studying models which exhibit infrared
slavery. 

An additional simplification occurs in the large $N_c$ limit
\beq
N_c\rightarrow \infty \quad \quad \quad \quad 
\beta^2\equiv \frac{g_s^2N_c}{2\pi}
\quad \mbox{fixed},
\eeq 
where sea quarks (more precisely, quark loops) are suppressed
and only planar diagrams survive.
Therefore, in a light-cone Fock space expansion, the valence
$\bar{q}q$ approximation for mesons becomes exact, and the
two body equation for the quark distribution amplitude
$\varphi_n(x)$ of mesons (the 't~Hooft equation \cite{thooft}) 
reads
\begin{eqnarray}
M^2_{n}\varphi_n(x) &=&
\left[
\frac{m_1^2 - \beta ^2}{x} + \frac{m_2^2 - \beta ^2}{1-x}
\right]
\varphi_n(x) - \beta ^2 \int_0^1{\rm d}y \,\frac{\varphi_n(y)}{(y - x)^2}\;
\nonumber\\
&\equiv&
\left[
\frac{m_1^2 - \beta ^2}{x} + \frac{m_2^2 - \beta ^2}{1-x}
\right]
\varphi_n(x) \;+\; V\, \varphi_n(x)\; .
\label{5}
\end{eqnarray}
Here $x$ 
denotes the light-cone momentum fraction carried by
the quark and $M_n$ is the invariant mass
of the meson. 

Eq.\,(\ref{5}) has a very physical interpretation: the light-cone
energy of the $q\bar{q}$ pair consists of a sum of the light-cone
kinetic energies of the quark and antiquark plus an interaction
term $V$.
The integral operator $V$ can be interpreted
as the momentum space representation of a linear potential.
The singularity of the
QCD-Coulomb interaction in Eq.\,(\ref{5}) is regularized using the principal 
value prescription,
with $\int_0^1 dy \frac{1}{(x-y)^2} = - \left(
\frac{1}{x}+\frac{1}{1-x}\right)$.

For practical purposes, it is convenient to rewrite
Eq.\,(\ref{5}) into the form where the singularity of the interaction
term is less severe:
\begin{equation}
M^2_{n}\varphi_n(x) =
\left(
\frac{m_1^2}{x} + \frac{m_2^2}{1\!-\!x}
\right)
\varphi_n(x) - \beta ^2 \int_0^1{\rm d}y \,
\frac{\varphi_n(y)\!-\!\varphi_n(x)}{(y - x)^2}\;.
\label{7}
\end{equation}
Note that Eq.\,(\ref{7}) can also be obtained as a variational
equation for the functional (Hamiltonian) defined by the quadratic
form
\begin{equation}
\matel{n}{{\cal H}}{n} =
\int_0^1 {\rm d}x\,
\left(
\frac{m_1^2}{x} + \frac{m_2^2}{1\!-\!x}
\right)
\varphi_n^2(x) + \frac{\beta^2}{2} \int_0^1 {\rm d}x\,{\rm d}y \,
\frac{(\varphi_n(y)\!-\!\varphi_n(x))^2}{(y - x)^2}\;.
\label{9}
\end{equation}
The interaction term in Eq.\,(\ref{9}) is non-negative and
the kinetic term is minimized for $\frac{x}{(1-x)} = \frac{m_1}{m_2}$,
i.e. when the ratio of momentum fractions carried by the quark and
antiquark equals their ratio of bare masses.
Therefore
the variational formulation Eq.\,(\ref{9}) of the 't~Hooft equation
yields a lower bound 
\begin{equation}
M_n \; > \; m_1+m_2
\label{11}
\end{equation}
(the inequality is saturated only for the massless pion where
$m_1=m_2=0$ and at $n=0$, with $\varphi_\pi(x)=1$). The mass of a bound
state exceeds the sum of the {\em bare} masses of the constituents, the
fact expected semiclassically.

A certain parity relation proves to be useful, it was first given in
Ref.\,\cite{callan} and reportedly ascends to 't~Hooft. We define
operator $K$ as
\begin{equation}
K\,\varphi(x)\;=\; \int_0^1{\rm d}y \, \frac{\varphi(y)}{y\!-\!x}\;.
\label{15}
\end{equation}
Then the following commutation relation holds:
\begin{equation}
[{\cal H}, K]\;=\;
\frac{m_1^2}{x} \int_0^1 \frac{{\rm d}y}{y} \,\varphi(y) \:-\:
\frac{m_2^2}{1\!-\!x} \int_0^1 \frac{{\rm d}y}{1\!-\!y} \,\varphi(y) \;.
\label{17}
\end{equation}
Since for eigenstates the expectation value of the commutator of
any operator $A$ with Hamiltonian vanishes,
\begin{equation}
\matel{n}{[{\cal H},
A]}{n}\;=\; 0\;,
\label{19}
\end{equation}
one has
\begin{equation}
0\;=\;\matel{n}{[{\cal H},
K]}{n} \;=\; m_1^2 \left(\int_0^1 \frac{{\rm d}y}{y}\,
\varphi_n(y)\right)^2- m_2^2 \left(\int_0^1 \frac{{\rm d}y}{1\!-\!y}\,
\varphi_n(y)\right)^2 \;.
\label{21}
\end{equation}
Therefore,
\begin{equation}
m_2 \int_0^1 \frac{{\rm d}y}{1\!-\!y}\,\varphi_n(y)
\;=\; -P_n\, m_1 \: \int_0^1 \frac{{\rm d}y}{y}\,\varphi_n(y)
\;,
\label{23}
\end{equation}
and $P_n$ is identified with the parity of the eigenstate $n$. This 
identification is
confirmed by comparing with explicit expressions for the matrix elements of 
the pseudoscalar and scalar densities
between the state $n$ and the vacuum \cite{callan} 
\begin{eqnarray}
\langle 0 | \bar{q}_2 q_1 | n \rangle
&=& \sqrt{\frac{N_c}{4\pi}}\:
\int_0^1 dx \left(\frac{m_1}{x}-\frac{m_2}{1\!-\!x}\right) \varphi_n(x)
\nonumber\\
\langle 0 | \bar{q}_2 i\gamma_5 q_1 | n \rangle
&=& \sqrt{\frac{N_c}{4\pi}}\:
\int_0^1 dx \left(\frac{m_1}{x}+\frac{m_2}{1\!-\!x}\right) \varphi_n(x) .
\end{eqnarray}
Using Eq.\,(\ref{23}), it is easy to see that
$\langle 0 | \bar{q}_2 q_1 | n \rangle \neq 0$ only for states
with $P_n=1$ and 
$\langle 0 | \bar{q}_2i\gamma_5 q_1 | n \rangle \neq 0$ only for states
with $P_n=-1$.

Integrating the 't~Hooft equation over $x$ and using the parity relation 
(\ref{23}) we get 
\bea
\nonumber
\int_0^1 \frac{{\rm d}x}{x}\, \varphi_n(x) & = &
\frac{M_n^2}{m_1(m_1\!-\!P_n m_2)} \int_0^1 \!{\rm d}x \,\varphi_n(x)\; ,
\\
\int_0^1 \frac{{\rm d}x}{1\!-\!x} \, \varphi_n(x) & = &
\frac{M_n^2}{m_2(m_2\!-\!P_n m_1)} \int_0^1 {\rm d}x \,\varphi_n(x)\; .
\label{24}
\eea
Similarly, integrating the 't~Hooft equation multiplied by $x$ yields 
\bea
\nonumber
\int_0^1 {\rm d}x \,x \varphi_n(x) \!\!&=&\!\!
\left[\frac{m_1^2\!-\!m_2^2}{M_n^2}+\frac{m_2}{m_2\!-\!P_{\!n} m_1}\right] 
\int_0^1 \!{\rm d}x \varphi_n(x) \!-\! 
\frac{\beta^2}{M_n^2} \int_0^1 
\!{\rm d}x \,\ln{\frac{1\!-\!x}{x}} \varphi_n(x) ,
\\
\nonumber
\int_0^1 \!{\rm d}x \,(1\!-\!x) \varphi_n(x) \!\!&=&\!\!
\left[\frac{m_2^2\!-\!m_1^2}{M_n^2}+\frac{m_1}{m_1\!-\!P_{\!n} m_2}\right] 
\int_0^1 {\rm d}x \varphi_n(x) \!+\! 
\frac{\beta^2}{M_n^2} \int_0^1 \!{\rm d}x \,\ln{\frac{1\!-\!x}{x}} 
\varphi_n(x) 
.
\label{24a}
\eea

Another useful relation for the 't~Hooft equation which holds for
arbitrary quark masses was derived by Burkardt \cite{mb:vt} and is
often referred to as the virial theorem. Let us denote
\begin{equation}
D\;=\;x \frac{\rm d}{{\rm d}x}\;, \qquad
\tilde D\;=\;(1\!-\!x) \frac{\rm d}{{\rm d}(1\!-\!x)}\;.
\label{25}
\end{equation}
A direct computation yields for the commutator of $D$ with the
interaction operator $V$ in the 't~Hooft equation the following result:
\begin{equation}
[D,V]\;=\; -V
\label{27}
\end{equation}
and, therefore,
\begin{equation}
[D,{\cal H}]\;=\; - {\cal H} + \frac{m_2^2-\beta^2}{(1-x)^2}\;.
\label{29}
\end{equation}
Likewise
\begin{equation}
[\tilde D,{\cal H}]\;=\; - {\cal H} + \frac{m_1^2-\beta^2}{x^2}\;.
\label{29a}
\end{equation}
Using again the fact that the expectation value of the commutator of 
any operator with  Hamiltonian vanishes in an energy eigenstate,
these two commutation relations yield the  relation which
holds for any eigenfunction:\footnote{As it stands, Eq.\,(\ref{31}) is
valid only for $m^2>\beta^2$. For $m^2\leq \beta^2$ it can be replaced by a 
subtracted version \cite{mb:vt}.}
\begin{equation}
M_n^2\;=\; (m_1^2\!-\!\beta^2)\aver{\frac{1}{x^2}} \:=\:
(m_2^2\!-\!\beta^2)\aver{\frac{1}{(1\!-\!x)^2}}\;.
\label{31}
\end{equation}

The 't~Hooft equations describing mesons with one heavy quark $Q$
acquire new symmetry which entails one to a number of additional
relations. In the rest of the paper we study these relations in the
static limit ($m_Q \to \infty$) and develop the $1/m_Q$ expansion around
this limit.

\section{Static version of the 't~Hooft equation}

In the heavy quark limit, i.e. when the mass of the quark
$m_1\rightarrow \infty$, the meson wave functions become sharply
peaked near $x\rightarrow 1$, since most of the momentum is then carried
by this quark.  Therefore, in order to study the
$m_1\equiv m_Q \rightarrow \infty$ limit, it is convenient to
introduce the 
nonrelativistic variables $M_n\!=\!m_Q+\tilde\epsilon_n$,
$\,t\!=\!(1\!-\!x)m_Q$ and
$\Psi_n(t)\!=\!\frac{1}{\sqrt{m_Q}}
\varphi_n\!\left(1\!-\!\frac{t}{m_Q}\right)$.
In these variables the 't~Hooft  equation (\ref{5}) takes the
form
\begin{equation}
\left(\tilde \epsilon_n \!+ \!
\frac{\tilde \epsilon_n^2\!+\!\beta^2}{2m_Q}\right) \Psi_n(t) \;=\;
\left(\frac{m_{\rm sp}^2\!-\!\beta^2}{2t} + \frac{t}{2}
\frac{1\!-\!\frac{\beta^2}{m_Q^2}}{1\!-\!\frac{t}{m_Q}}
\right) \Psi_n(t)\;-\;
\frac{\beta^2}{2}\, \int_{0}^{m_Q}
{\rm d}s\:\frac{\Psi_n(s)}{(t\!-\!s)^2}\;\;.
\label{41}
\end{equation}
We assume that in the notations of the previous section $m_1\!=\!m_Q$ is
large, and $m_{\rm sp}\!=\!m_2$ will be denoted simply as $m$ below.
The limit $m_Q \!\gg \! \beta$ is obtained by expanding the second
term in the r.h.s.\ of Eq.\,(\ref{41}) in $t/m_Q$ and extending it to the
interval $[0, \infty)$:\footnote{Extending the interval to $[0, \infty)$
introduces errors that are only ${\cal O}(1/m_Q^4)$.}
$$
\left(\tilde\epsilon_n \!+ 
\!\frac{\tilde\epsilon_n^2\!+\!\beta^2}{2m_Q}\right)
\Psi_n(t) \:=\:
\left(\frac{m^2\!-\!\beta^2}{2t} \right) \Psi_n(t)+
\left(1\!-\!\frac{\beta^2}{m_Q^2}\right) \left[\frac{t}{2} +
\frac{t^2}{2m_Q}+ \frac{t^3}{2m_Q^2} + \ldots \right] \Psi_n(t)
$$
\begin{equation}
\!\!\!\! -\;
\frac{\beta^2}{2}\, \int_{0}^{\infty}
{\rm d}s\:\frac{\Psi_n(s)}{(t-s)^2}\;\;.
\label{43}
\end{equation}
Performing $1/m_Q$ expansion, it is convenient to study the eigenvalues
\begin{equation}
\epsilon_n \equiv \tilde\epsilon_n \!+
\!\frac{\tilde\epsilon_n^2\!+\!\beta^2}{2m_Q}
\label{44}
\end{equation}
of the equation themselves rather than directly $\tilde\epsilon_n$. This will
be assumed later when we study $1/m_Q$ corrections to the static limit.
(The explicit factor $1\!-\!\beta^2/m_Q^2$ can be eliminated by
properly rescaling $t$.)

The static limit is obtained neglecting all terms suppressed by inverse
powers of $m_Q$ \cite{burkswan,burk,D2}:
\begin{equation}
\epsilon_n \Psi_n(t) \;=\;
\frac{m^2\!-\!\beta^2}{2t} \Psi_n(t) +
\frac{t}{2} \Psi_n(t)
\;-\;
\frac{\beta^2}{2}\, \int_{0}^{\infty}
{\rm d}s\:\frac{\Psi_n(s)}{(t\!-\!s)^2}\;\;.
\label{45}
\end{equation}
This is a stationary Schr\"{o}dinger-type equation for the
one-dimensional system with the static Hamiltonian
\begin{equation}
{\cal H} \;=\;  {\cal H}_{\rm stat} \;=\;
\frac{m^2\!-\!\beta^2}{2t} + \frac{t}{2}
\;-\; \frac{\beta^2}{2}\, \int_{0}^{\infty}
\:\frac{{\rm d}s}{(t\!-\!s)^2}\;\;.
\label{47}
\end{equation}
Similar to Eq.\,(\ref{9}), the expectation value of the Hamiltonian over
a state $\Psi_n$ can be written in the form
\begin{equation}
\matel{n}{{\cal H}_{\rm stat}}{n} =
\int_0^\infty {\rm d}t\,
\left(
\frac{m^2}{2t} + \frac{t}{2}
\right)
\Psi_n^2(t) + \frac{\beta^2}{4} \int_0^\infty {\rm d}s\,{\rm d}t \,
\frac{(\Psi_n(s)-\Psi_n(t))^2}{(s \!-\! t)^2}\;.
\label{49}
\end{equation}
Since $\frac{m^2}{2t} + \frac{t}{2} \ge m$, Eq.\,(\ref{49}) suggests
that $\epsilon_n >m$. This lower bound follows also directly from
the general lower bound Eq.\,(\ref{11}).

The large-$t$ asymptotics of the static eigenfunctions is obtained
directly:
\begin{equation}
\Psi_n(t) \; \stackrel{t \to \infty}{\longrightarrow} \;
\frac{\beta^2 F^{(n)}}{t^3}\;,
\label{51}
\end{equation}
where
\begin{equation}
F^{(n)} \,\equiv \, \int_0^\infty {\rm d}t \,\Psi_n(t) \;=\;
 \lim_{m_Q\to\infty} \sqrt{\frac{\pi m_Q}{N_c}}\, f_n\;,
\label{52}
\end{equation}
and $f_n $ is the usual annihilation constant of a meson
\cite{callan}. The combination
$F^{(n)}$ has finite large-$N_c$ and large-$m_Q$ limits. Together with
various moments (integrals of $\Psi^2_n(t)$ with powers of $t$) it
plays an important role in the heavy quark expansion. For future use,
it is convenient to define a set of operators acting on $\Psi(t)$,
\begin{equation}
J_l \Psi\,=\, \int_0^\infty {\rm d}t \,t^l \Psi(t) \;, \qquad
J_0 \equiv J\;,
\label{55}
\end{equation}
and matrix elements
\begin{equation}
F_l^{(n)} \,= \, \int_0^\infty {\rm d}t \, t^l \Psi_n(t) \;.
\label{55b}
\end{equation}
Then, for example, $\left(F^{(n)}\right)^2 = \matel{n}{J_0}{n}$.
Of course, the integrals in Eqs.\,(\ref{55}) and (\ref{55b}) literally 
converge only for $-1 \le l <2 $.

The static analogue of the operator $K$ in Eq.\,(\ref{15}) takes the
form
\begin{equation}
K\, \Psi(t) \,=\, \int_0^\infty {\rm d}s \, \frac {\Psi(s)}{s\!-\!t} \;,
\label{57}
\end{equation}
and its commutator with the Hamiltonian
\begin{equation}
[H, K]\,=\,\frac{m^2}{2t} \int_0^\infty \frac{{\rm d}s}{s}\;-\;
\frac{1}{2} \int_0^\infty {\rm d}s \;=\;
\frac{m^2}{2t} J_{-1} -\frac{1}{2} J_0\;.
\label{58}
\end{equation}
For any energy eigenstate $n$, the equation $\matel{n}{[H,K]}{n}=0$ holds, 
and we thus find
\begin{equation}
m^2 \left|\int_0^\infty \frac{{\rm d}t}{t}\, \Psi_n(t)\right|^2\;=\;
\left|\int_0^\infty {\rm d}t\, \Psi_n(t)\right|^2\;,
\label{59}
\end{equation}
or
\begin{equation}
m \int_0^\infty \frac{{\rm d}t}{t}\, \Psi_n(t)\;=\; -P_n\,
\int_0^\infty {\rm d}t\, \Psi_n(t)\;,
\label{61}
\end{equation}
i.e.\ $mF^{(n)}_{-1}\!=\!-P_n F^{(n)}$.
Hereafter we call states for which 
$P_n\!=\!-1$ and $P_n\!=\!1\;$ $P$-odd and $P$-even, respectively. 
Of course, Eq.\,(\ref{61}) can also be obtained directly by applying the heavy
quark limit to the finite mass parity relation Eq.\,(\ref{23}),
which is the reason why corrections to the static limit do not 
spoil the parity classification.

Another relation among the moments of the wavefunction can be obtained by
integrating the static equation (\ref{45}) from $0$ to $\infty$,
yielding
$$
2 \epsilon_n F^{(n)}\,=\, m^2 F_{-1}^{(n)}\ + F_{1}^{(n)}\;,
$$
which with the help of the above parity relation can be written as
\begin{equation}
F_{1}^{(n)}\;=\;
\left(2 \epsilon_n \!+\! m P_n\right) F^{(n)}\;.
\label{63}
\end{equation}

Non-diagonal matrix elements of $K$ are also expressed in terms of
$\epsilon$ and $F$:
\begin{equation}
\matel{n}{K}{l}\;=\; \frac{P_n P_l \!-\!1}{2}
\frac{1}{\epsilon_n\!-\!\epsilon_l} \, F^{(n)} F^{(l)}\;.
\label{65}
\end{equation}
Certain useful relations emerge also from the commutation relation
\begin{equation}
[{\cal H}, t] \;=\; - \frac{\beta^2 K}{2}\;,
\label{66}
\end{equation}
and the obvious commutator $[t,K]=-J_0$. Thus, for example, 
$[t,[t,{\cal H}]]=-\frac{\beta^2}{2}J_0$ and hence 
$\matel{n}{[t,[t,{\cal H}]]}{n}= -\frac{\beta^2}{2} 
\left(F^{(n)}\right)^2\,$.\,\footnote{ 
This relation is the so-called fourth sum rule for the Darwin operator.} 

In the following, we will derive a tower of relations among the moments of
$\Psi_n^2$. For this purpose, we consider the operators
\begin{equation}
D_n\,=\, t^n \frac{{\rm d}}{{\rm d}t}\;, \qquad
D_1\,=\, t \frac{{\rm d}}{{\rm d}t}\,\equiv D\;.
\label{67}
\end{equation}
A direct computation yields
\begin{equation}
[D_n, {\cal H}] \,=\, - n t^{n-1} {\cal H} + \frac{n\!-\!1}{2}
(m^2\!-\!\beta^2)
t^{n-2} + \frac{n\!+\!1}{2}  t^n - \frac{\beta^2}{2} \sum_{k=0}^{n-3}\,
(k\!+\!1)(n\!-\!k\!-\!2) t^{n\!-\!k\!-\!3} J_k \, .
\label{69}
\end{equation}
For $n=0,1,2$ the last sum is absent. The first few relations take the
form
\begin{equation}
\begin{array}{rcll}
\left[\frac{{\rm d}}{{\rm d}t}, {\cal H}\right] & \; = \;
& - \frac{m^2\!-\!\beta^2}{2t^2} + \frac{1}{2} & n=0 \\

\left[t \frac{{\rm d}}{{\rm d}t}, {\cal H}\right] & \;=\; & - {\cal H}+t &
n=1 \\

\left[t^2\frac{{\rm d}}{{\rm d}t}, {\cal H}\right] & \;=\; & -
 2t\,{\cal H} + \frac{1}{2} (m^2\!-\!\beta^2) + \frac{3}{2} t^2 & n=2 \\

\left[t^3 \frac{{\rm d}}{{\rm d}t}, {\cal H}\right] & \;=\;
 & - 3t^2\,{\cal H} + (m^2\!-\!\beta^2)t + 2t^3  - \frac{\beta^2}{2} J_0 & n=3 \\

\left[t^4\frac{{\rm d}}{{\rm d}t}, {\cal H}\right] & \;=\;
& - 4t^3\,{\cal H} + \frac{3}{2} (m^2\!-\!\beta^2)t^2 + \frac{5}{2} t^4
- \beta^2 t J_0 - \beta^2 J_1  & n=4 \;.
\end{array}
\label{71}
\end{equation}

Taking the expectation values of the operator relations
in Eqs.\,(\ref{71}) we obtain the moments $\aver{t^n}$
in terms of the bound-state energies and decay constants $F$:
\begin{equation}
\begin{array}{lrcl}
n=1 \qquad  & \aver{t} &=& \epsilon \\
n=2 \qquad  & 3\aver{t^2} &=&  4\aver{t}^2 -(m^2\!-\!\beta^2) \\
n=3 \qquad  & 4\aver{t^3} &=&
6\aver{t^2}\aver{t}-2(m^2\!-\!\beta^2)\aver{t}+ \beta^2 F_0^2 \\
n=4 \qquad  & 5\aver{t^4} &=&
8\aver{t^3}\aver{t}-3(m^2\!-\!\beta^2)\aver{t^2}+ 4 \beta^2 F_0 F_1 \;.
\end{array}
\label{75}
\end{equation}
Note that $n\ge 5$ does not literally lead to meaningful 
relations since they would involve
divergent terms.
The case $n=0$ yields the relation
\begin{equation}
(m^2\!-\!\beta^2)\, \int_0^\infty \frac{{\rm d}t}{t^2}\, \Psi^2(t)\;=\;1\;,
\label{77}
\end{equation}
i.e. the virial theorem Eq.\,(\ref{31}) for the light quark  in the limit 
where  the other quark is static.
As we have mentioned above, it is literally valid at $m > \beta$, for smaller 
$m$ it can 
be understood, for example, as an analytic continuation in the mass of
light antiquark or in a subtracted form.
As we will show later, these moments are important
in the $1/m_Q$ expansion since the leading $1/m_Q$ corrections
in Eq.\,(\ref{43}) are simple powers of $t$.

The relation $\aver{t} = \epsilon_n$ is also a direct consequence of the virial
equation (\ref{31}) (the one which involves $m_Q^2$) expanded to the
first nontrivial order in $1/m_Q$. It was first derived in
Ref.\,\cite{burk} (see also \cite{D2}). In QCD the bound-state
energies $\epsilon_n$ are usually denoted by $\La_n$.

The next important parameter of the heavy quark expansion is the
kinetic expectation value
$\mu_\pi^2=\frac{1}{2M_{H_Q}} \matel{H_Q}{\bar{Q} (i\vec{D}\,)^2
Q}{H_Q}$. 
In the limit $m_Q\to\infty$ the
operator of the spacelike momentum takes the simple form, since the
$Z$-graph contributions can be neglected, the light-cone combination of
momentum is given by $x M_{H_Q}$ and the time component can be excluded
using the equation of motion $D_0\, Q=m_Q\, Q $. 
In the 't~Hooft model, in the limit $m_Q\to\infty$ 
this immediately leads to 
\begin{equation}
\mu_\pi^2\,=\, \int_0^\infty 
{\rm d}t\, (t\!-\!\bar{t}\,)^2\, \Psi^2(t)\;=\;
\aver{t^2}-\aver{t}^2\;.
\label{81}
\end{equation}
Combining the $n\!=\!2$ relation in Eq.\,(\ref{75}) with the virial equation
($n\!=\!1$) one finds 
\begin{equation}
3\mu_\pi^2\;=\; \La^2-m^2+\beta^2,\qquad\qquad \La\equiv \epsilon_n\;,
\label{83}
\end{equation}
for any bound state.

In the absence of actual chromomagnetic field in $D\!=\!2$ the next
operator is represented by the Darwin term
\begin{equation}
\rho_D^3\,=\,\frac{1}{2M_{H_Q}} \matel{H_Q}{\bar{Q}
(\mbox{\small$-\frac{1}{2}$} 
\vec{D} \vec{E}\,) Q}{H_Q} \;=\; \frac{\beta^2 F^2}{4}
\label{84}
\end{equation}
(the last relation is obtained using the equation of motion for the
gauge field, and factorization valid at $N_c \!\to\! \infty$). At the same
order a nonlocal zero-momentum correlator of the kinetic operators
appears as well, which will be addressed in the next section.

Before proceeding to the IW functions, let us mention an upper bound on
the decay constants $F^{(n)}$. It is obtained using one of the
Sobolev's inequalities bounding the $L_{\infty}$ norm {\it via} $L_2$
and $L_2^1$ norms in one dimension:
\begin{equation}
\left| f(a)\right| \,\le\,  2^{1/2} \left[ \int {\rm d}z\,
\left| f(z)\right|^2
\right]^{1/4} \,
\left[ \int {\rm d}z\, \left| f'(z)\right|^2
\right]^{1/4} \qquad \mbox{for any $a$.}
\label{91}
\end{equation}
In terms of the Fourier transform of $f(z)$, $\Psi(t)$ it takes the
form
\begin{equation}
\left|\int {\rm d}t\, \Psi(t)\, \right|  \,\le\,  \pi^{1/2}
\left[ \int {\rm d}t\, \Psi^2(t)\, \right]^{1/4} \,
\left[ \int {\rm d}t\, t^2\,\Psi^2(t)\, \right]^{1/4} \;.
\label{93}
\end{equation}
Applied to the static wavefunction it reads
$$
F\;\le\;  \sqrt{\pi}\left(\La^2+ \mu_\pi^2\right)^{1/4}\: = \:
\sqrt{\pi}\left(\frac{4\La^2\!-\!m^2\!+\!\beta^2}{3}\right)^{1/4}\;,
$$
or
\begin{equation}
\rho_D^3 \,< \frac{\pi}{4}\: \beta^2
\left(\La^2 \!+\! \mu_\pi^2\right)^{1/2}
\;.
\label{95}
\end{equation}
These bounds are a $D\!=\!2$ counterpart of the bounds discussed in QCD
in Ref.\,\cite{boost}.

Since the inequality in Eq.\,(\ref{93}) is saturated only by 
functions of the form
$1/(c+t^2)$, a solution of the 't~Hooft equation cannot saturate the
bound (\ref{95}). It is interesting, however, that for the ground
states with light spectator quarks, $m\lsim \beta$ the decay constant
$F$ numerically almost saturates the bound, within only a few percent.

The operator $t$ plays a special role for the static equation
(\ref{45}): the first and the last terms in ${\cal H}_{\rm stat}$ are
homogeneous functionals of rank $-1$ with respect to $t$, while the term
$\propto t/2$ has rank $+1$. It breaks the dilatation invariance of the
eigenstate problem. This operator is the analogue of the operator
${\cal D}$ of Ref.\,\cite{optical} representing the part of the full
 trace of the energy-momentum tensor $\theta_{\mu\mu}$ 
associated with the light
degrees of freedom (Sect.\, II, Eq.\,(9)). Likewise, there are many
relations for various observables in the 't~Hooft model, involving
the operator $t$. Here we consider matrix elements of $t$.

Using the commutator Eq.\,(\ref{66}) we write
$$
\matel{k}{t}{n}\,=\,\frac{1}{\epsilon_k\!-\!\epsilon_n}
\matel{k}{[{\cal H},t]}{n} \,=\,-
\frac{\beta^2}{2(\epsilon_k\!-\!\epsilon_n)} \matel{k}{K}{n}= -
\frac{\beta^2}{2(\epsilon_k\!-\!\epsilon_n)^2} \matel{k}{[{\cal
H},K]}{n} \;=
$$
\begin{equation}
=\;
- \frac{\beta^2}{4(\epsilon_k\!-\!\epsilon_n)^2}
\left(m^2 F^{(k)}_{-1} F^{(n)}_{-1} - F^{(k)}_{0} F^{(n)}_{0} \right)
\,=\,
\frac{\beta^2}{4(\epsilon_k\!-\!\epsilon_n)^2} \, F^{(k)}_{0}
F^{(n)}_{0}\, \left(1- P_k P_n \right)\;,
\label{101}
\end{equation}
where relations (\ref{58}) and (\ref{61}) have also been used. This can
be cast into the form
\begin{equation}
\matel{k}{t\!-\!{\cal H}}{n} \; =\;
\frac{\beta^2}{2(\epsilon_k\!-\!\epsilon_n)^2} \,
F^{(k)}_{0} F^{(n)}_{0}\, \left(\frac{1- (-1)^{k-n}}{2} \right)
\label{103}
\end{equation}
which embeds both $k\!=\!n$ and $k\!\ne\! n$ (we
have used the fact that $P_k P_n = (-1)^{k-n}$). The above equation
shows that the operator $t\!-\!{\cal H}$ is $P$-odd, i.e. its matrix
elements do not vanish  only between the states of opposite
parity. This makes sense since, in the static limit, 
$t\!-\!{\cal H}$ is simply $\bar{Q}iD_z Q$.

Using the second of the commutation relations (\ref{71}) we have for the
non-diagonal matrix elements of the dilatation operator $D$
\begin{equation}
\matel{k}{t\frac{{\tiny \rm d}}{{\rm \small d}t}}{n} \;=\;
- \frac{\beta^2}{2(\epsilon_k\!-\!\epsilon_n)^3} \,
F^{(k)}_{0} F^{(n)}_{0}
\cdot \left\{
\begin{array}{ll}
0 & \mbox{wrong parity}\\
1 & \mbox{right parity}
\end{array}
\right.
\;.
\label{105}
\end{equation}
These matrix elements determine the so-called {\it oscillator
strengths} -- the Small Velocity (SV) transition amplitudes between the
heavy quark states (usually the ground and the ``$P$-wave'' states
in actual QCD). Eq.\,(\ref{105}) allows one to prove an important
symmetry relation for the IW function.

\subsection{IW function}

The IW function determines the transition amplitudes between two
heavy-quark states induced by a current bilinear in two heavy quark
fields. In the present context it can be defined as a diagonal scalar
current in the heavy quark limit $m_Q \to \infty$:
\begin{equation}
\frac{1}{2\sqrt{p'_0 p_0}} \matel{H_Q^{(k)}(p')}{\bar{Q}Q}{H_Q^{(n)}(p)}
\;=\; \xi_{nk}
\left(\frac{(pp')}{M_{H_Q}^2}\right)\;,
\qquad
\frac{(pp')}{M_{H_Q}^2}=
(vv')\equiv w\;.
\label{111}
\end{equation}
In the 't~Hooft model the IW functions are given by the following
expression in terms of the static wavefunctions:
$$ \xi_{nk}\;=\;
\frac{2}{1+w\pm\sqrt{w^2-1}}\, \int_0^\infty {\rm d}t\:
\Psi_k\left(t\right) \Psi_n\left([w \mp \sqrt{w^2-1}]t\right)
\;=
$$
\begin{equation}
=\;
\frac{2\sqrt{z}}{1+z} \, \int_0^\infty {\rm d}t\:
\Psi_{k}\left(\frac{t}{\sqrt{z}}\right)
\Psi_{n}\left(\sqrt{z} t \right)
\;,
\label{113}
\end{equation}
where
\begin{equation}
w = \frac{1+z^2}{2z}\,, \qquad z=w\pm \sqrt{w^2-1}
\;.
\label{C4}
\end{equation}
The expression for the IW function was obtained in
Refs.\,\cite{burkswan} and \cite{D2WA}.

Let us note that each value of $w\ne 1$ can be represented by two
different values of $z$ corresponding to two possible values of the
square root in Eq.\,(\ref{113}), such that $z_1z_2=1$. They must yield
the same value of $\xi$, up to a sign:
\begin{equation}
\xi_{nk}(z)\;=\; P_n P_k \: \xi_{nk}(1/z)\;,
\label{115}
\end{equation}
which, for $n \!\ne\! k$ looks like a miraculous property of the 't~Hooft
equation \cite{burkswan}.
Alternatively, the above property can be written as
\begin{equation}
\xi_{nk}(w)\;=\; P_n P_k \: \xi_{kn}(w)\;.
\label{116}
\end{equation}
Now we can demonstrate it explicitly.\footnote{We thank R.\ Lebed for
its cross checks in the numerical computations.}

Using the fact that $D\equiv t\frac{{\rm d}}{{\rm d}t}$ is the 
generator of scale 
transformations, i.e. 
\begin{equation}
e^{\ln (a)D} f(t) \equiv e^{\ln (a) t\frac{{\rm d}}{{\rm d}t} } f(t) = 
f(at)
\end{equation}
for an arbitrary function $f(t)$,
the IW function can be written in the form
\begin{equation}
\xi_{nk}(z)\;=\; \frac{2\sqrt{z}}{1\!+\!z}\:
\matel{k}{{\rm e}\,^{(D+\frac{1}{2})\ln{z}}}{n}\;,
\qquad D=t \frac{{\rm d}}{{\rm d}t}\;.
\label{117}
\end{equation}
Note that the operator $D+\frac{1}{2}$ is antihermitean (antisymmetric),
so that $\|{\rm e}\,^{(D+\frac{1}{2})\ln z}\state{n}\|$ $= \|
\state{n}\|$. This ensures the so-called first sum rule expressing the
unit probability of the transition to arbitrary final state in the heavy
quark limit (in the SV approximation it is known as the Bjorken sum
rule).

To calculate the diagonal matrix elements of $D$ one can use the
identity
\begin{equation}
\int {\rm d}t\; \Psi'(t)\,\Psi(t)\:f(t)\:=\:
\frac{1}{2} \int {\rm d}t\,
\left[
\frac{{\rm d}}{{\rm d}t} \left( \Psi^2(t)\:f(t) \right)
\,-\,
\Psi^2(t) \:f'(t)
\right]
\,=\:
-\frac{1}{2}\,\aver{f'(t)}
\label{121}
\end{equation}
valid for arbitrary $f(t)$. In particular, it shows that
the expectation values of
$D+ \frac{1}{2}$ vanish. Together with relation (\ref{105}) we see
that only  even powers of $D+ \frac{1}{2}$ survive in the exponent
in Eq.\,(\ref{117}) when $\state{n}$ and $\state{k}$ have the same
parity, and only odd powers contribute if the parity of the two states
is  opposite:
\begin{equation}
\xi_{nk}\;=\; \frac{2\sqrt{z}}{1\!+\!z}
\cdot \left\{
\begin{array}{cl}
\matel{k}{\cosh{[(D+\frac{1}{2})\ln{z}]}}{n} \;\; &
n\!-\!k=\mbox{even}\\
\matel{k}{\sinh{[(D+\frac{1}{2})\ln{z}]}}{n} \;\; &
n\!-\!k=\mbox{odd}
\end{array}
\right.  \;.
\label{123}
\end{equation}
This proves the symmetry properties Eqs.\,(\ref{115}),\,\,(\ref{116}) and
ensures that the IW functions are analytic at $vv'=1$, in spite of the
branch point in $z$ as a function of $vv'$.

\subsection{SV sum rules}

Important constraints on the transition amplitudes between heavy
flavor hadrons  and on the parameters of the heavy quark expansion
follow from the sum rules, in particular, in the small velocity (SV)
limit. In 1+1 dimensions the first four sum rules in the heavy quark
limit take the form
\bea
\nonumber
\rho_k^2-\frac{1}{4} &=& \sum_l
\tau_{lk}^2\\ 
\mbox{\small$\frac{1}{2}$} \: \epsilon_k &=&  
\sum_l (\epsilon_l\!-\!\epsilon_k) \tau_{lk}^2\\
\nonumber
\left(\mu_\pi^2\right)_k &=& \sum_k (\epsilon_l\!-\!\epsilon_k)^2
\tau_{lk}^2\\ 
\left(\rho_D^3\right)_k &=& \sum_l
(\epsilon_l\!-\!\epsilon_k)^3 \tau_{lk}^2\;.
\label{A2}
\eea
The so-called `oscillator strengths'
$\tau$ parameterize the transition amplitudes into the opposite-parity
states in the SV limit, and  $\rho^2$ denotes 
the slope of the elastic IW function:
\beq
\frac{1}{2M} \matel{l}{\bar{Q}\gamma_\mu Q}{k}\;=\;  \tau_{lk}
\varepsilon_{\mu \nu}v^\nu\;,
\qquad
\frac{1}{2M} \matel{k(\vec{v})}{\bar{Q}\gamma_0 Q}{k(0)}\;=\; 
 1-\rho^2_{k} \frac{\vec{v}^{\,2}}{2}\, +\, 
{\cal O}\left(\vec{v}^{\,4}\right)
\label{A4}
\eeq
($\vec v$ is the velocity of the final state hadron). Therefore,
\beq
\tau_{lk} = \matel{l}{\,t\frac{{\rm d}}{{\rm d}t}\!+\!\frac{1}{2}\,}{k}\;,
\qquad \qquad
\rho^2_{k}-\frac{1}{4} = 
\matel{k}{\left(t\frac{{\rm d}}{{\rm d}t}\!+\!\frac{1}{2}\right)^2}{k}\;.
\label{A6}
\eeq
The first sum rule then becomes obvious being a consequence of 
completeness of the  eigenstates. Other sum rules are straightforward as
well.

Consider, for example the third sum rule for the kinetic
operator. Using the commutator with $n=1$ in Eq.\,(\ref{71}), we have
\beq
\sum_l \tau_{lk}^2\,(\epsilon_l\!-\!\epsilon_k)^2\;=\; 
\sum_l \matel{k}{t\!-\!{\cal H}}{l}  \matel{l}{t\!-\!{\cal H}}{k} \,=\,
\matel{k}{t^2}{k}-\left(\matel{k}{t}{k}\right)^2\,=\,
\left(\mu_\pi^2\right)_k\;.
\label{A8}
\eeq
Similarly, we get for the second, ``optical'' sum rule
\beq
\sum_l \tau_{lk}^2\,(\epsilon_l\!-\!\epsilon_k)\;=\; 
-\sum_l \matel{k}{t}{l}  \matel{l}{\,t\frac{{\rm d}}{{\rm d}t}
\!+\!\frac{1}{2}\,}{k} 
\,=\,
-\matel{k}{\,t^2\frac{{\rm d}}{{\rm d}t}\!+\!\frac{t}{2}\,}{k}\,=\,
\frac{\epsilon_k}{2}\;.
\label{A10}
\eeq

The fourth sum rule for the Darwin operator can be directly obtained
by inserting the complete set of states into the commutator
$[t,[t,{\cal H}]]=-\frac{\beta^2}{2}J_0\,$:
\beq
\sum_l \tau_{lk}^2\,(\epsilon_l\!-\!\epsilon_k)^3\;=\;-\frac{1}{2}
\matel{k}{\,[t,[t,{\cal H}]]\,}{k} \;=\; \frac{\beta^2 (F^{(k)})^2}{4} = 
\left(\rho_D^3\right)_k\;.
\label{A12}
\eeq

\section{$1/m_Q$ expansion}

In practice, it is often necessary to account for the first few
$1/m_Q^k$ corrections to the static limit $m_Q\!\to\!\infty$, since in
actuality these effects are non-negligible not only for charm, but even
for beauty hadrons. In studies of the 't~Hooft model there appears an
additional motivation: the available numerical
approaches often apply only to the finite quark masses. The solution of
the static equation (\ref{45}) is approximated by the solution of the
generic finite-mass 't~Hooft equation where $m_Q$ is taken large but
finite.  For computational reasons $m_Q$ cannot be taken too large, and
control over the `spurious' $1/m_Q$ corrections becomes mandatory even
for studies of the pure static case.

In this section we will study the  leading $1/m_Q$ 
corrections to the axial decay constant, meson masses and the
kinetic energy of the heavy quark.
The $1/m_Q$ expansion is carried out by applying to Eq.\,(\ref{43}) the
standard formalism of non-covariant time-independent perturbation
theory used in QM.
Since the leading $1/m_Q$ corrections in Eq.\,(\ref{43}) 
involve only powers of $t$, it is possible
to derive exact expressions for the first few terms in the 
$1/m_Q$ expansion for these observables
that depend only on the moments of the static
wavefunction and of the structure function. Using the results from
Sect.~3, one can then express the coefficients
in the $1/m_Q$ expansion solely in terms of the static binding
energy and the decay constant.

Although we will perform the $1/m_Q$ expansion using 
old-fashioned time-ordered perturbation theory, we will
introduce here the corresponding notations which 
resemble those used in field theory, where $1/m_Q$
corrections to various expectation values are given by 
correlators of the type 
\begin{equation}
-\matel{k}{\int_{-\infty}^{\infty} {\rm d}\tau\: iT\{A(0), \delta{\cal
H}(\tau)\}}{k}\;.
\label{131}
\end{equation}
Heisenberg operators
$O(\tau)$ are understood as ${\rm e}^{i{\cal H}\tau}\, O(0)\, {\rm
e}^{-i{\cal H}\tau}$; we assume that the Schr\"{o}dinger operators we
deal with, do not depend on $\tau$ explicitly.

We then denote for the stationary problem 
\begin{eqnarray}
\matel{k}{iT\{ A, B \}}{k}\;&\equiv&\;
\matel{k}{\int_{-\infty}^{\infty} {\rm d}\tau\: iT\{A(0), B(\tau)\}}{k}
\nonumber\\
\;&=&\;\sum_{n\ne k}
\frac{\matel{k}{A}{n}\matel{n}{B}{k}}{\epsilon_n-\epsilon_k} \,+\,
\sum_{n\ne k}
\frac{\matel{k}{B}{n}\matel{n}{A}{k}}{\epsilon_n-\epsilon_k}\;.
\label{133}
\end{eqnarray}
The similar expectation value can be defined for the
time-ordered product of arbitrary number of operators. We consider
$$
{\cal H} \to {\cal H} + \alpha A + \beta B + \gamma C +...
$$  
and put 
\begin{equation}
\matel{k}{iT\{ A, B, C,... \}}{k}\;\equiv\;
\left.-\left(\frac{\partial}{\partial\alpha}\frac{\partial}{\partial\beta}
\frac{\partial}{\partial\gamma}... \;\epsilon_k(\alpha,\beta,
\gamma,...)
\right)\right|_{\alpha=\beta=\gamma=...=0}\;.
\label{131a}
\end{equation}

Two basic relations hold for such $T$-products of two operators:
\begin{equation}
\matel{k}{iT\{[{\cal H},A], B \}}{k}\;=\;  \matel{k}{[A,B]}{k}
\label{135}
\end{equation}
and
\begin{eqnarray}
\nonumber
\matel{k}{iT\{A{\cal H}, B \}}{k}&\; =\;&
\epsilon_k \,\matel{k}{iT\{A, B \}}{k} - \matel{k}{AB}{k} +
\matel{k}{A}{k}\matel{k}{B}{k}\;,\\
\matel{k}{iT\{{\cal H}A, B \}}{k}&\; =\;&
\epsilon_k \,\matel{k}{iT\{A, B \}}{k} - \matel{k}{BA}{k} +
\matel{k}{A}{k}\matel{k}{B}{k}\;.
\label{137}
\end{eqnarray}
The first relation has a transparent meaning: since $[{\cal H},A]= -i
\frac{{\rm d}A}{{\rm d}t}$, \,Eq.\,(\ref{135}) is a form of integrating
by parts:
\begin{equation}
\matel{k}{\int {\rm d}\tau\: T\left\{\frac{{\rm d}A(\tau)}{{\rm d}\tau},
B(0)\right\}}{k}
=\matel{k}{\int {\rm d}\tau\: \frac{{\rm d}}{{\rm d}\tau}\,
T\{A(\tau), B(0)\}}{k} + \matel{k}{[A(0),B(0)]}{k}\;.
\label{139}
\end{equation}
The obvious relation
$$
\matel{k}{iT\{{\cal H}^l, A \}}{k}\;=\; 0
$$
holds as well, which will be used later.

In the static limit, $\aver{t}$ equals to the bound-state energy
$\epsilon$. It is often necessary to know how $\aver{t}$ changes under
various perturbations $\delta {\cal H}$. The answer is readily obtained
using Eqs.\,(\ref{71}) and (\ref{135}):
\begin{equation}
\delta \, t_{kk} \;=\; \matel{k}{iT\{ t, \delta {\cal H} \}}{k} \,=\,
-\matel{k}{[t\frac{{\rm d}}{{\rm d}t}, \delta {\cal H}]}{k}\;.
\label{143}
\end{equation}
Since $D_1=t\frac{{\rm d}}{{\rm d}t}$ generates scale transformations
in $t$, one finds
\begin{equation}
[t\frac{{\rm d}}{{\rm d}t}, O ] \;=\; \mbox{Dim}[O]\cdot O\;.
\label{145}
\end{equation}
Say, $[t\frac{{\rm d}}{{\rm d}t}, t^l]= l\, t^l$. Therefore, any
perturbation which is a homogeneous rank-$l$ functional of $t$, satisfies
\cite{D2WA}
\begin{equation}
\delta\, \aver{t}\; = \; -l \,\aver{\delta{\cal H}_l}\;.
\label{147}
\end{equation}
For example, for $\delta {\cal H}=\lambda t$ one finds
$\delta \langle t\rangle=-\lambda$, a result that can
be easily verified by direct evaluation, since the
exact result reads $\langle t\rangle_\lambda
=\langle t \rangle/\sqrt{1+2\lambda}$.

The same property holds for non-local operators as well. For example,
\begin{equation}
\matel{k}{iT\{t, A, B \}}{k}\;=\; -\left(D[A]\!+\!D[B]\!+\!1\right)\,
\matel{k}{iT\{A, B \}}{k}\;,
\label{149}
\end{equation}
or
$$
\matel{k}{iT\{(t\!-\!2{\cal H}), A, B \}}{k}\;=\; 
-\left(D[A]\!+\!D[B]\!-\!1\right)\,
\matel{k}{iT\{A, B \}}{k}\;,
$$
where $D[A]$, $D[B]$ denote the $t$-dimension of operators $A$ and $B$.

The above properties parallel the relations for the operator of the
trace of the energy-momentum tensor for the light degrees of freedom in actual
QCD discussed in Ref.\,\cite{optical}. This similarity was elucidated
in the previous section.

The $m_Q$-suppressed terms in the r.h.s.\ of Eq.\,(\ref{43}) playing  the
role of the perturbation $\delta{\cal H}$ are given by $t^2/2m_Q$,
$\,t^3/2m_Q^2$, {\it etc.} Therefore, in the $1/m_Q$ expansion one
typically needs to compute $T$-products Eq.\,(\ref{133}) with
the operator $t^2$ (or $t^3$, in higher order). As exemplified above,
this can be done directly using the relations in 
Eqs.\,(\ref{135}) and (\ref{137}):
\beq
\label{153}
\matel{k}{iT\{t^2, A \}}{k}  =  -\frac{2}{3} \matel{k}{[D_2, A]}{k} -
\frac{4}{3}\epsilon_k \matel{k}{[D_1, A]}{k} -
\frac{4}{3}\matel{k}{t A}{k}+
\frac{4}{3}\epsilon_k \matel{k}{A}{k}\;,
\eeq
$$
\matel{k}{iT\{t^3 \!-\! \frac{\beta^2 J_0}{4} , A \}}{k} = 
-\frac{1}{2} \matel{k}{[D_3, A]}{k} \!-\!\epsilon_k \matel{k}{[D_2, A]}{k}
\!-\!\frac{1}{2} (4\epsilon_k^2\!-\!m^2\!+\!\beta^2) \matel{k}{[D_1, A]}{k}
$$
\beq
-\; \frac{3}{2} \matel{k}{t^2 A}{k} \!-\!2 \epsilon_k \matel{k}{t A}{k}
+\frac{1}{2} (8\epsilon_k^2\!-\!m^2\!+\!\beta^2) \matel{k}{A}{k}
\;,
\label{154}
\eeq
{\it etc.}

As an application of these relations, we obtain
\begin{equation}
\matel{k}{iT\{t^2, t^2\}}{k}\;=\; -\frac{4}{3} \left(
2\aver{t^3}+\aver{t}\aver{t^2}\right)\,=\,
-\left[\frac{64}{9} \epsilon_k^3 \!-\!
\frac{28}{9}\epsilon_k (m^2\!-\!\beta^2)+
\frac{2}{3} \beta^2 F_0^2 \right]
\;.
\label{155}
\end{equation}
This correlator governs the $1/m_Q$ corrections to the average
$\aver{t^2}$ which, in turn, determines the kinetic expectation value
in the static limit. Likewise
\begin{equation}
\matel{k}{iT\{t^2, t^l\}}{k}\;=\;
- \frac{2l+4}{3} \matel{k}{t^{l+1} }{k}-
\frac{4}{3}(l\!-\!1) \epsilon_k \matel{k}{t^l }{k}
\;.
\label{157}
\end{equation}

Similarly we get the analytic expression for the $1/m_Q$ corrections
to the decay constants $F^{(k)}$. Indeed,
$(F^{(k)})^2=\matel{k}{J_0}{k}$, and
$$
\delta_{1/m_Q}\,(F^{(k)})^2 \; = \;\frac{1}{2m_Q}
\matel{k}{iT\{t^2, J_0\}}{k} \;=
$$
$$
\frac{1}{2m_Q}
\matel{k}{-\frac{2}{3}[D_2,J_0]}{k} + \frac{1}{2m_Q}\, \frac{4}{3}
\left\{\epsilon_k \matel{k}{iT\{t, J_0\}}{k} -\matel{k}{t J_0}{k} +
\matel{k}{t}{k}\matel{k}{J_0}{k} \right\} =
$$
\begin{equation}
\frac{1}{2m_Q} \left\{ -\frac{4}{3} F_0^{(k)}F_1^{(k)}
\!-\!\frac{4}{3} \epsilon_k \matel{k}{[D_1, J_0]}{k} \!-\!
\frac{4}{3} F_0^{(k)}F_1^{(k)} \!+\! 
\frac{4}{3}\epsilon_k (F_0^{(k)})^2 \right\}
=
- \frac{1}{2m_Q} \frac{8}{3} F_0^{(k)}F_1^{(k)} ,
\label{159}
\end{equation}
where we have used that $[D_1,J_0]=J_0 $ and $[D_2,J_0]=2J_1$. Thus,
\begin{equation}
\frac{\delta_{1/m_Q}\,F^{(k)}}{F^{(k)}} = -\frac{1}{m_Q} \frac{2}{3} \,
\frac{F_1^{(k)}}{F_0^{(k)}}\,=\,
-\frac{2(2\epsilon_k \!+\! m P_k)}{3m_Q} \;.
\label{161}
\end{equation}
In the last equation we used relation Eq.\,(\ref{63}) for
$F_1^{(k)}$.  The $1/m_Q$ corrections to $F$ turn out significant
(c.f.\ Refs.\,\cite{burkswan,mende}).

It is often advantageous to define the axial decay constant {\it via}
the pseudoscalar current rather than the axial current:  
\begin{equation}
\frac{1}{2 M_{H_Q}} \matel{0}{\bar{q} i\gamma_5 Q}{k}
\;=\;\frac{1}{2}\,\tilde{f}^{(k)}\;,
\label{163}
\end{equation}
then
\begin{equation}
\tilde{f}^{(k)}\,=\, f^{(k)}\,\frac{M_k}{m_Q \!+\! m}
\qquad \mbox{and}
\qquad
\tilde{F}^{(k)}\,=\, F^{(k)}\,\frac{M_k}{m_Q\!+\!m}\;.
\label{164}
\end{equation}
Similar to what is observed in actual QCD, the $1/m_Q$ corrections to
the ground-state $\tilde{F}$ are smaller,
\begin{equation}
\frac{\delta_{1/m_Q}\,\tilde{F}^{(k)}}{\tilde{F}^{(k)}}\;=\;
-\frac{\epsilon_k \!+\! m (3\!+\!2P_k)}{3m_Q}\;.
\label{165}
\end{equation}

The analytic expression (\ref{161}) agrees with the numerical
computations performed in Ref.\,\cite{burkswan} for the ground state.
In terms of $c_k$ introduced there to quantify these preasymptotic
corrections,
\begin{equation}
c_k\,=\, \frac{5}{6} \epsilon_k + m P_k\;.
\label{167}
\end{equation}
In the nonrelativistic case $\epsilon_k \to m$ holds, and for the
negative-parity ground state one has the correct limit $c=-m/6$
\cite{burkswan}.  
For the first excitation, however, one would have $c_1
\to 11m/6$.

It is not difficult to derive the expression for the $1/m_Q$
correction to the light-cone wavefunction itself generated by 
perturbation $\frac{t^2}{2m_Q}$:
\beq
\Psi_k(t)\;=\; \Psi_k^{\infty}(t) + \frac{1}{m_Q} \Psi_k^{(1)}(t)\;.
\label{B2}
\eeq
Using the commutators in Eqs.\,(\ref{71}) we get
\beq
\Psi_k^{(1)}(t)\;=\; \frac{1}{3} \left(
t^2\frac{{\rm d}}{{\rm d}t} + 2\epsilon_k t\frac{{\rm d}}{{\rm d}t} 
+ 2\epsilon_k \right) \Psi_k^{\infty}(t)\;.
\label{B4}
\eeq
Note that simply replacing $m_Q$ by $M_{H_Q}\!\simeq\! m_Q \!+\! \epsilon$
when passing from $\varphi_k(x)$ to $\Psi_k(t)$, would amount to
adding only 
$\frac{1}{m_Q}\epsilon_k\left(t\frac{{\rm d}}{{\rm d}t}+
\frac{1}{2}\right)\Psi_k(t)$. 
The remaining part 
$\frac{1}{3m_Q}\left(t^2\frac{{\rm d}}{{\rm d}t}-
\epsilon_k t\frac{{\rm d}}{{\rm d}t}+
\frac{\epsilon_k}{2}\right)$
actually changes the shape.

Using similar techniques, it is straightforward to obtain explicit
$1/m_Q$ expansion of the meson mass in the 't~Hooft model; we consider
it through order $\beta^3/m_Q^2$ corresponding to the order discussed
in case of QCD \cite{optical}. A straightforward evaluation of the
expectation value of Eq.\,(\ref{43}) yields (see also \cite{D2WA})
\begin{equation}
M_{\!H_Q}\!-\!m_Q\,=\, \aver{t}_{\infty} +
\frac{\aver{t^2}_{\infty}\!-\! \aver{t}_{\infty}^2\!-\!\beta^2}{2m_Q} +
\frac{4\aver{t^3}\!+\!4 \aver{t}^3 \!+\! \aver{iT\{t^2\!,t^2\}}
\!-\! 4\aver{t}\aver{t^2}}{8m_Q^2}
\,+\,{\cal O}\!\left(\frac{\beta^4}{m_Q^3}\right)
.
\label{171}
\end{equation}
Here all expectation values correspond to the static limit
$m_Q\to\infty$, {\it i.e.} do not implicitly depend on $m_Q$. Using the
previously derived relations, especially Eqs.\,(\ref{75}) and
(\ref{155}), all terms can be expressed only in terms of the
static binding energy $\La$ and the axial decay constant of the
respective state.

Finally, the object usually appearing in the $1/m_Q$ expansion of the
diagonal matrix elements to this order, is the zero-momentum correlator
of operators $\vec{\pi}^2=\bar{Q}(i\vec{D}\,)^2 Q$ which represent the
$1/m_Q$ piece of the Hamiltonian,
\begin{equation}
-\rho_{\pi\pi}^3\;=\;
\frac{1}{2} \, \matel{k}{iT\{\vec{\pi}^2 \!,\vec{\pi}^2\}}{k}\;.
\label{173}
\end{equation}
In particular, it determines the $1/m_Q$ variation of the kinetic
expectation value $\matel{k}{\bar{Q}(i\vec{D}\,)^2 Q}{k}$ itself in
the actual finite-$m_Q$ hadron. The expression for
$\rho_{\pi\pi}^3$ in the 't~Hooft model is most simply obtained using
the above mentioned relation  $\bar{Q}(-iD_z) Q = t\!-\!{\cal H}$ which holds
for the zero-momentum matrix elements in the static limit. Therefore, 
we simply need to compute $\matel{k}{iT\{(t\!-\!{\cal H})^2,
(t\!-\!{\cal H})^2\}}{k}$.
In this way we obtain
\begin{equation}
-2 \rho_{\pi\pi}^3\;=\; \matel{k}{iT\{\vec{\pi}^2,\vec{\pi}^2\}}{k}\;=\;
4\aver{t^3}+  4\aver{t}^3 + \aver{iT\{t^2,t^2\}} - 4\aver{t} \aver{t^2}
- \frac{\beta^2}{2} F^2 \;=
\label{187}
\end{equation}
$$
=\; -\left( \frac{4}{9} \epsilon_k^3 - \frac{4}{9} \epsilon_k
(m^2\!-\!\beta^2) + \frac{1}{6} \beta^2 F^2 \right)
\;=\; -\left(\frac{4}{3}\mu_\pi^2\epsilon_k  + \frac{1}{6} \beta^2 F^2 \right)
\;.
$$
This correlator is numerically large.

We can compare the
$1/m_Q$ expansions discussed above with the general operator expansion
Ref.\,\cite{optical,D2}
valid in arbitrary gauge theory:
\begin{equation}
M_{H_Q}\!-\!m_Q\,=\, \La +
\frac{\mu_\pi^2\!-\!\beta^2}{2m_Q}  -
\frac{1}{m_Q^2} \left[-\frac{1}{8} \aver{iT\{\vec\pi^2,\vec\pi^2\}} +
\frac{\rho_D^3}{4} \right] + ...\;, \qquad \rho_D^3 = \frac{\beta^2
F^2}{4}\,.
\label{181}
\end{equation}
Similarly -- and even a bit simpler -- is to consider the expansion of
the scalar density which is given precisely by \cite{D2}
\begin{equation}
\frac{1}{2M_{H_Q}}\,
\aver{\bar{Q}Q}\; = \; \frac{m_Q}{M_{H_Q}}\, \aver{\frac{1}{x}}
\label{183}
\end{equation}
and, on the other hand, use the similar OPE for $\aver{\bar{Q}Q}$:
\begin{equation}
\frac{1}{2M_{H_Q}}\, \matel{H_Q}{\bar{Q}Q}{H_Q}\; = \;
1\,-\,\frac{\aver{\bar{Q}(i\vec{D}\,)^2 Q}  \!-\!\beta^2}{2m_Q^2} -
\frac{\rho_D^3}{2m_Q^3}\,+\,{\cal O}\!\left(\frac{\beta^4}{m_Q^3}\right)
\,.
\label{185}
\end{equation}
The explicit computations show that these equations are satisfied
with $\rho^3_{\pi\pi}$ given by Eq.\,(\ref{187}).

It is interesting that it is possible to derive a closed expression
for the expectation value of the `kinetic' operator
$\bar{Q}(iD_z)^2Q$ in terms of the 't~Hooft wavefunction for
arbitrary mass $m_Q$:
$$
\frac{1}{2M_{H_Q}}\matel{H_Q}{\bar{Q}(iD_z)^2Q}{H_Q} \; = \qquad\qquad\qquad
\qquad\qquad\qquad
$$
\beq 
\frac{m_Q}{2M_{H_Q}} \left[
M_{H_Q}^2\aver{x}-m_Q^2\aver{\frac{1}{x}} +
\frac{\beta^2}{4} \left(\frac{M_{H_Q}^2}{m_Q(m_Q\!-\!m P)} 
\int_0^1 {\rm d}x\, \varphi(x) \right)^2\,
\right]
\label{E4}
\eeq
(this expression assumes a certain ultraviolet regularization of the
operator, see below), where $P$ is parity of $H_Q$. The idea is the following.

In the rest frame of the meson the expectation value of 
$\bar{Q}[(iD_0)^2 \!+\! (iD_z)^2]Q$ is simply  
$2\aver{\bar{Q}(iD_-)^2Q}$ and, therefore 
\beq
\frac{1}{2M_{H_Q}}
\matel{H_Q}{\bar{Q}[(iD_0)^2\!+\!(iD_z)^2]Q}{H_Q} = 
m_Q M_{H_Q} \int_0^1 {\rm d}x\, x \varphi^2(x)\;. 
\label{E6}
\eeq

The complementary combination 
of momentum operators $\bar{Q}[(iD_0)^2\!-\!(iD_z)^2]Q$ can be determined 
using the general identity
\begin{equation}
\int d^D x\; \bar{Q} Q(x) \;=\;
\int d^D x \; \left\{ \bar{Q}\gamma_0 Q \;+\;
\bar{Q}\frac{(iD_0\!-\!m_Q)^2 -(i\vec{D}\,)^2 + \frac{i}{2}\sigma^{\mu\nu}
G_{\mu\nu}}{2m_Q^2} Q
\right\}
\label{E8}
\end{equation}
valid in arbitrary dimension. In $D=2$ one has 
$\frac{i}{2}\sigma^{\mu\nu} G_{\mu\nu} = \tilde G i\gamma^5$ with
$\tilde G=\frac{1}{2}\epsilon^{\mu\nu} G_{\mu\nu}=G_{01}$. Since
$\aver{\bar{Q}iD_z Q}=\aver{\bar{Q}iD_0 iD_z Q} =0$ and the
light-cone combination 
$\aver{\bar{Q}(iD_0\!-\!iD_z) Q}$ is again directly expressed {\it via} the
't~Hooft wavefunction, Eq.\,(\ref{E8}) yields the necessary equation
for $\aver{\bar{Q}[(iD_0)^2\!-\!(iD_z)^2]Q}$ involving, however the
expectation value $\aver{\bar{Q}\tilde G i\gamma^5 Q}$.

The gluonic field strength $G_{01}$ can be explicitly written {\it via} quark
current in the light-cone gauge: 
\beq
G_{01}= -g_s^2 \frac{1}{\partial_-} J^+
\label{E10}
\eeq
which leads to a non-local four-fermion operator. We note, however,
that the current $J^+$ includes not only the spectator
quark $q$, but also $\bar{Q} \gamma^+ \frac{\lambda^a}{2}Q$. This term
described by the bare loop leads to the ultraviolet divergent
expression. We simply discard this contribution in $J^+$ in
Eq.\,(\ref{E10}), which fixes a certain renormalization
procedure. Then we get 
$$
\aver{\bar{Q} \frac{i}{2}\sigma^{\mu\nu} G_{\mu\nu} Q}=
\aver{\bar{Q}\tilde G i\gamma^5 Q} =  
\frac{\beta^2}{4} \frac{m_Q}{M_{H_Q}} 
\int_0^1 {\rm d}x \,\varphi(x) \int_0^1 {\rm d}y \,\varphi(y)
\frac{1}{x\!-\!y} \left(\frac{1}{x}\!-\! \frac{1}{y} \right)
$$
\beq
\qquad\qquad\qquad = \; -\frac{\beta^2}{4} \frac{m_Q}{M_{H_Q}} 
\left(\int_0^1 \frac{{\rm d}x}{x} \,\varphi(x)\right)^2\;.
\label{E12}
\eeq
This finally yields Eq.\,(\ref{E4}) where we have used
Eqs.\,(\ref{24}) 
to express the last term {\it via} the decay constant of the meson.

A note of caution must be voiced regarding this derivation, however. 
In the way described above 
we obtain the {\em bare} operator $\bar{Q}(iD_z)^2 Q$. 
It does include a finite contribution from the domain of 
momenta of order $m_Q$ even in the leading order in $m_Q$. On the
contrary, in the heavy quark expansion we are interested only in the
physics originating from momenta essentially below the scale of the
heavy quark mass itself. The expressions Eqs.\,(\ref{173}),
(\ref{187}) refer just to such low-energy effective
operator. Therefore, in general the literal comparison of the two
expectation values is not too instructive. 
It is easy to check that already to the leading order in $m_Q$ the
two expressions differ by the amount $\frac{\beta^2}{2}$ attributed
to the domain of momenta $\sim m_Q$. 

Here we note an interesting feature of the exact expectation value of
the local quark-gluon operator $\bar{Q}\tilde G i\gamma_5 Q$ in
Eq.\,(\ref{E12}). At $m_Q \to 0$ it has an $1/m_Q$ singularity
regardless of the mass of the second quark in the meson: $\int_0^1
{\rm d}x/x \,\varphi(x) \propto 1/m_Q$ at $m_Q \ll \beta$. Yet we know
that no appropriate massless {\it physical} states exist in the model 
at $m_q\ne 0$ (the ground $Q\bar Q$ state has negative parity, and
their pairs are $1/N_c$ suppressed). The singularity technically
emerges due to massless gluon propagator, however gluon is absent
from the physical spectrum in $D=2$.\footnote{We are grateful to
A.\,Vainshtein for informing us of existing examples of similar IR
singularities in physical amplitudes in the absence of contributing
massless particles, in low dimensions. Reportedly, such a situation
is excluded in $D > 3$.}

It must be noted, however, that  careful treatment of passing to
the light-cone coordinates in the computations of the similar {\it
vacuum} expectation value $\matel{0}{\bar{\psi}\tilde G i\gamma_5
\psi}{0}$ in the 't~Hooft model yielded additional terms which
canceled the $1/m_\psi$ pole observed in Ref.\,\cite{zhit} and led to 
a finite result at $m_\psi \to 0$. A possibility of similar
subtleties in the computation of the meson expectation values deserves 
further studies. We are grateful to A.\,Zhitnitsky for pointing out
and discussing this problem.

\subsection{$1/m_Q^2$ correction at zero recoil}

As another application of the $1/m_Q$ expansion, we briefly consider
here the second-order nonperturbative corrections to the zero-recoil
$B\to D^{(*)}$ transition amplitude. At this kinematic point the
deviation from the elastic IW function (which is unity here) appears
at the level $1/m_{c,b}^2\,$, which provides a method of extracting
$|V_{cb}|$. The corrections, however, are shaped by strong dynamics
at the typical hadronic scale and at present cannot be evaluated from 
the first principles. The existing estimates, in particular for the
axial $B\to D^*$ amplitude, rely on the sum rules derived in 
Refs.\,\cite{optical,vcb}:
\beq
|F_{D^*}|^2 +
\sum_{k\ne 0}|F_k|^2
\; = \;
\xi_A \;-\;
\frac{\mu_G^2}{3m_c^2} -
\frac{\mu_\pi^2\!-\!\mu_G^2}{4}
\left(\frac{1}{m_c^2}+\frac{1}{m_b^2}+\frac{2}{3m_cm_b}
\right)
\;+\;
{\cal O}\left(\frac{1}{m_Q^3}\right) \,,
\label{C2}
\eeq
where $F_k$ are the zero-recoil transition amplitudes to the excited
states, $F_k\propto 1/m_Q$, $\xi_A$ is the short-distance
renormalization factor, and $\mu_\pi^2$, $\mu_G^2$ are expectation
values of the kinetic and chromomagnetic operators,
respectively. One then has \cite{vcb}
\beq
F_{D^*}\;=\;\xi_A^{1/2} -  \left[\frac{\mu_G^2}{6m_c^2} +
\frac{\mu_\pi^2\!-\!\mu_G^2}{8}
\left(\frac{1}{m_c^2}+\frac{1}{m_b^2}+\frac{2}{3m_cm_b}
\right)\right]
\cdot (1+\chi) \;+\;
{\cal O}\left(\frac{1}{m_Q^3}\right) \,,
\label{C3}
\eeq
where a positive quantity $\chi$ parameterizes the magnitude of the
sum of the excitation probabilities in the l.h.s.\ of the sum rules,
in terms of the local operator term in the r.h.s.\ of Eq.\,(\ref{C2}):
\beq
\sum_{k\ne 0}|F_k|^2 \;=\; \chi \cdot \left[\frac{\mu_G^2}{3m_c^2} +
\frac{\mu_\pi^2\!-\!\mu_G^2}{4}
\left(\frac{1}{m_c^2}+\frac{1}{m_b^2}+\frac{2}{3m_cm_b}
\right)\right]\;.
\label{C6}
\eeq
The expressions for the excitation amplitudes were elaborated in
Ref.\,\cite{optical}.
Following Ref.\,\cite{vcb}, existing numerical estimates of $F_{D^*}$ assume 
(somewhat arbitrarily) that $\chi$ can vary up to $1$, that is,
$\chi=0.5\pm 0.5$. 

We computed $\chi$ analytically in the 't~Hooft
model. Since spin and chromomagnetic field are absent in two
dimensions, only the kinetic operator acts here. The sum rule takes
the form 
\beq
F_{D}^2 +
\sum_{k\ne 0}|F_k|^2
\; = \;
\xi_A \;-\;
\left(\frac{1}{m_c}-\frac{1}{m_b}\right)^2
\frac{\mu_\pi^2}{4}
\;+\;
{\cal O}\left(\frac{1}{m_Q^3}\right) \,,
\label{C8}
\eeq
and the excitation amplitudes $F_k$ to the leading order are given by 
\beq
F_k \;=\; \frac{1}{2}\left(\frac{1}{m_c}-\frac{1}{m_b}\right) 
\frac{\matel{k}{\bar{Q}\pi_z^2Q}{0} }{\epsilon_k-\epsilon_0}
\;.
\label{C10}
\eeq
Similarly, $\chi$ is defined through
\beq
\sum_{k\ne 0}|F_k|^2 \; = \; \chi \cdot 
\left(\frac{1}{m_c}-\frac{1}{m_b}\right)^2 \frac{\mu_\pi^2}{4}
\;.
\label{C12}
\eeq

The sum of $F_k^2$ can be computed using the same technique as was
elaborated in the previous sections. However, 
the corrections to the amplitudes we consider are not expressed
anymore {\it via} only positive integer moments, and include
expectation values of operators with higher derivatives. Yet they can 
be expressed in terms of the slope of the IW function $\rho^2$. We
give here the final result
\beq
\chi \;=\; \frac{10}{21} +\frac{5}{63}\frac{\epsilon_0^2}{\mu_\pi^2}- 
\frac{4}{21}\left(\rho^2\!-\!\frac{1}{4}\right)\,=\,
\frac{5}{7} + 
\frac{5}{21}\frac{m^2\!-\!\beta^2}{\epsilon_0^2\!-\!m^2\!+\!\beta^2} 
- \frac{4}{21}\left(\rho^2\!-\!\frac{1}{4}\right)\;.
\label{C14}
\eeq

For light spectator quark the value of $\chi$ turns out to be about
$0.55$. This is surprisingly close to the central value guestimated
in Ref.\,\cite{vcb} for the case of actual QCD.

\section{Summary and Outlook}

We have studied the 't~Hooft model in the $m_Q \!\rightarrow \!\infty$
limit. Our main result are exact relations for the
heavy quark kinetic energy, as well as 
moments of the 't~Hooft wavefunction and of the structure function
in this limit, which allow to express these observables
in terms of only the heavy quark binding energy $\bar{\Lambda}$
and the axial decay constant $F^k$.

In the 't~Hooft model, these moments appear as coefficients
in the $1/m_Q$ expansion for various observables.
As an example, we calculated coefficients in the $1/m_Q$ 
expansion of meson masses, decay constants and heavy quark
kinetic energies. Using the above relations, we were
able to express the corresponding $1/m_Q$ coefficients in 
terms of $\bar{\Lambda}$ and $F^k$. 

Likewise, we derived the expressions for the oscillator strengths and 
verified a set of the SV sum rules in the heavy quark limit. As an
application of the developed $1/m_Q$ expansion, we computed the
nonperturbative $1/m_Q^2$ corrections to the zero recoil $B\to D$
transition amplitude. 

Although the 't~Hooft model is in principle
``numerically solvable'',
many observables can only be determined with very limited 
precision in practical calculations. This is particularly
the case for observables in the limit where one of the
quarks becomes heavy. In this
case the 't~Hooft wavefunction becomes extremely asymmetric
and many numerical techniques, which are otherwise rather 
powerful for finite quark masses, fail to produce numerically 
reliable results. For this regime, where the heavy quarks
are not infinitely heavy, it is often advantageous 
to perform the $1/m_Q$ expansion beyond the leading
order. The corresponding expansion coefficients that we derived 
involve only properties of the 't~Hooft wavefunctions in the 
static limit. Moreover, through the use of exact
relations, the expansion coefficients can be expressed in
terms of $m_Q\rightarrow \infty $ properties of the wavefunctions 
that can be calculated numerically with sufficiently high
accuracy.

The developed analytic $1/m_Q$ expansion allows to carry out 
precision studies, in the framework of the 't~Hooft model, of such an 
intriguing and poorly understood phenomenon as violation of local
duality in heavy flavor decays \cite{lebur}. The question of its
magnitude has a particular phenomenological significance in the
domain of moderately heavy quarks, where reliability of the
asymptotic expansions is unknown {\it a priori}, and numerical
computations are unavoidable.

The developed technique can be used to test, on the example of the
't~Hooft model, various approximations routinely applied to the
actual beauty decays. Some of them will be reported in
Ref.\,\cite{lebur}, while others further deserve dedicated studies.

\vspace*{.2cm} 

{\bf Acknowledgments:}~~This work has been supported by the NSF 
under grant number PHY96-05080, 
by the DOE under grant number DE-FG03-95ER40965, 
by NATO under the reference PST.CLG 974745, 
by RFFI grant \#\,99-02-18355, and in part by TJNAF.  
N.U. thanks I.\,Bigi, R.\,Lebed, M.\,Voloshin
and A.\,Zhitnitsky for helpful comments and M.\,Shifman and
A.\,Vainshtein for encouraging interest. N.U.\ gratefully
acknowledges the hospitality of Physics Department of the Technion
and the support of the Lady Davis grant during completion of this
paper.

\end{document}